\documentclass[12pt]{article}
\usepackage{amssymb,latexsym,epsfig,cite}

\newcommand{\be}{\begin{equation}}
\newcommand{\ee}{\end{equation}}
\def\bea{\begin{eqnarray}}
\def\eea{\end{eqnarray}}
\begin{document}

\thispagestyle{empty}

\begin{center}
{\Large \bf Position Dependent Mass Oscillators and
Coherent\\[1ex]
States}
\end{center}

\vskip1cm

\begin{center}
Sara Cruz~y~Cruz$^{1,2}$ and Oscar Rosas-Ortiz$^{1}$\\[2ex]
{\footnotesize $^{1}$\it Departamento de F\'{\i}sica, Cinvestav,
AP 14-740, 07000 M\'exico~DF, Mexico}\\
{\footnotesize $^{2}$\it Secci\'on de Estudios de Posgrado e
Investigaci\'on, UPIITA-IPN, Av. IPN 2508, CP 07340 M\'exico~DF,
Mexico}
\end{center}

\begin{abstract}
\noindent The solving of the Schr\"odinger equation for a
position-dependent mass quantum system is studied in two ways.
First, it is found the interaction which must be applied on a mass
$m(x)$ in order to supply it with a particular spectrum of
energies. Second, given a specific potential $V(x)$ acting on the
mass $m(x)$, the related spectrum is found. The method of solution
is applied to a wide class of position-dependent mass oscillators
and the corresponding coherent states are constructed. The
analytical expressions of such position-dependent mass coherent
states preserve the functional structure of the Glauber states.
\end{abstract}

\section{Introduction}

The problem of calculating the energies of a quantum system
endowed with position-dependent mass $m(x)$ and subjected to a
given interaction represents an interface between theoretical and
applied physics. Its antecedent can be identified with the concept
of effective mass, introduced in the forties to discuss the motion
of electrons or holes in semiconductors \cite{Pec46}. Successfully
applied in describing the formation of shallow energy levels due
to impurities in crystals, the effective mass theory was strongly
developed in the fifties \cite{Kit54}. Further insights were given
in the calculation of superlattice band structures for which the
band edges and the masses are position dependent. In such context,
it was stressed that the correct effective Hamiltonian consists of
the kinetic term $\frac{1}{4} \{ P^2, \frac{1}{m(x)} \}$ instead
of the conventional expression $\frac{P^2}{2m}$ \cite{Bas81}. That
is, the Hermiticity of the Hamiltonian is a part of the problem if
the mass is not a constant.

The subject has embraced potentials other than the periodic ones
over the years. Indeed, the energy bands and periodic-like
interactions appearing quite naturally in semiconductor physics
are substituted with point spectra and properly defined potentials
in mathematical physics
\cite{Lev95,Alh02,Des00,Car04,Gan06,Bag06,Sch06,Sch07,Mus06,
Des08,Mil97,Pla99,Gon02,Bag08,Koc03,Que04,Cru07,Roy05}. This new
perspective has inspired intense activity in the looking for new
exactly solvable potentials in Quantum Mechanics. Of particular
interest, Susy-QM (factorization or intertwining method)
\cite{Coo01,Mie04,And04,Don07} and group-theory \cite{Wey55,Mil68}
(see also \cite{Neg00b}) represent the most fruitful approaches on
the matter \cite{Mil97,Pla99,Gon02,Bag08,Koc03,Que04,Cru07,Roy05}.
However, the literature focuses on mainly one of the two sides of
the problem. Namely, in order to get a particular spectrum, the
appropriate mass-function $m(x)$ and potential $V(x)$ are usually
looked for. A more deeper insight is necessary if one is
interested on a particular mass $m(x)$ subject to a previously
defined interaction rather than looking for the recovering of a
specific spectrum.

In this work we analyze the two sides of the position-dependent
mass problem by following the transformation scheme of the
Schr\"odinger equation reported in \cite{Neg00b}. In a first step
the equation involving $m(x)$ is mapped to the equation of a
constant mass $m_0$. After obtaining some general results, we
study the eigenvalue problem connected with diverse
position-dependent mass oscillators. In general, we distinguish
between two fundamental kinds of oscillators. The first one is
characterized by exhibiting the conventional set of equidistant
energies $\hbar \omega_0 (n+1/2)$, no matter the explicit form of
$m(x)$ or $V(x)$. The oscillators of the second kind, on the other
hand, are endowed with position-dependent mass $m(x)$ and
subjected to the conventional oscillator interaction $V_{\rm
osc}(x) = m_0 \omega_0^2 x^2/2$. The spectra of these last
oscillators depend on the explicit form of the mass-function. In
this way, we are able to compare the behavior of a quantum system
of mass $m(x)$ with that of a particle of mass $m_0$ when both of
them are acted by the same oscillator-like potential. One of our
motivations to analyze such oscillators is due to the fact that,
as far as we know, there is a lack of results including the
coherent states for position-dependent mass systems.

Originally derived for electromagnetic fields \cite{Gla06}, the
features of the standard coherent states (Glauber states) are a
consequence of the oscillator dynamical algebra \cite{Per86}. They
are usually constructed as eigenstates of the annihilation
operator but are shown to minimize the uncertainty relation
between position and momentum as well. A third property is that
the Glauber states are displaced versions of the ground
wavefunction. For other systems, generalized coherent states (CS)
can be constructed through algebraic techniques (see e.g.
\cite{Per86,Zha90}). In general, the CS show not all the three
basic properties of the Glauber states. They have been recently
studied in connection with non-linear Susy-algebras
\cite{Fer07,Bar08} (see also the reviews \cite{Fer05,Dod02}),
classical motion models for the P\"oschl-Teller potential
\cite{Cru08}, anharmonic vibrations in diatomic molecules
\cite{Ang08}, Landau levels \cite{Ali08} and the Penning trap
\cite{Fer09}. With the present work we introduce some families of
position-dependent mass oscillator coherent states.

The paper is organized as follows. In Section~\ref{sec2} the
Schr\"odinger equation of a position-dependent mass system is
connected with the equation of a constant mass $m_0$. The
solutions are interrelated by a mapping for which the Hamiltonians
are isospectral. The main difficulty is that the Hamiltonian of
the mass $m_0$ includes an effective potential which, in general,
makes the related equation as complicated to solve as the initial
one. Here, the problem will be faced by either selecting the
appropriate mass-function $m(x)$ or by fixing the order in which
$m$ is entangled with $P$ in the Hamiltonian. In
Section~\ref{sec3} the previous general results are particularized
to the harmonic oscillator potential. As a first result, it is
shown that potentials behaving as a confining odd-root-law,
$\ln^2$, or $\sinh^2$ give rise to the quantum oscillator
energies. The singular oscillator $V=x^2+\alpha x^{-2}$ is
analyzed as a special case. On the other hand, it is also shown
that the acting of an oscillator potential on $m(x)$ involves the
energies of a constant mass $m_0$ subject to either a confining
even-power-law or the $\sinh^2$-like potentials. To deal with
these last potentials it will be unavoidable the numerical
approximation of the solutions.

In Section~\ref{sec4} a position-dependent mass Hamiltonian is
shown to be factorized by a couple of two mutually adjoint
operators, the commutator of which depends on the explicit form of
$m(x)$. The initial Hamiltonian is then intertwined with a new one
in such a way that they are isospectral. Factorization operators
can be properly selected to work as ladders when acting on the
eigenfunctions derived in Section~\ref{sec3}. The related coherent
states are constructed as eigenfunctions of the annihilation
operator. These position-dependent mass CS are shown to have the
same analytical form as the Glauber states. Moreover, they are
also displaced versions of the ground state and minimize the
uncertainty relation between $P$ and $X$. Finally, in the very
last section of the paper some concluding remarks are given.

\section{The eigenvalue equation}
\label{sec2}

Let us consider the one-dimensional Hamiltonian
\be
H_a = \frac12 m^a P m^{2b} P m^a + V \equiv K_a +V, \qquad 2a+2b
=-1
\label{hamil1}
\ee
where the mass $m>0$ and the potential $V$ are functions of the
position, $K_a$ is the kinetic term of $H_a$ and $P$ fulfills
$[X,P]=i\hbar$, with $X$ the position operator. We shall use
${\cal D}^{(a)}$ to represent the domain of definition of $H_a$,
i.e. ${\cal D}^{(a)} \equiv Dom (H_a)$.

In the position representation $X=x$ and $P = -i\hbar
\frac{d}{dx}$, so we have:
\be
[f(x),P] =i\hbar f'(x), \qquad '\equiv \frac{d}{dx}.
\label{deriva}
\ee
This last commutator allows us to express the Hamiltonian
(\ref{hamil1}) as follows
\be
\begin{array}{rl}
H_a & = \displaystyle \frac{1}{2m} P^2 + \frac{i\hbar}{2m} \left(
\frac{m'}{m} \right) P -\frac{\hbar^2}{2m} \left( \frac{a}{m^2}
\right) \left[
mm'' -(2+a) (m')^2 \right] +V\\[2ex]
& = \displaystyle \alpha_1 \frac{d^2}{dx^2} + \alpha_2
\frac{d}{dx} + \alpha_3,
\end{array}
\label{hamil2}
\ee
with
\be
\alpha_1= -\frac{\hbar^2}{2m}, \quad \alpha_2 = -\alpha_1 \left(
\frac{m'}{m} \right) = \alpha_1', \quad \alpha_3 = \alpha_1 \left(
\frac{a}{m^2} \right) \left[mm''- (2+a)(m')^2 \right] +V.
\label{alfas}
\ee
In order to solve the eigenvalue equation
\be
H_a \psi(x) = E\psi(x)
\label{eigen1}
\ee
we first rewrite the functions $\psi$ to read
\be
\psi(x) = e^{g(x)} \varphi(x),
\label{map1}
\ee
with $g$ and $\varphi$ two functions to be determined and such
that
\be
\int_{{\cal D}^{(a)}} \vert \psi(x) \vert^2 dx = \int_{{\cal
D}^{(a)}} \vert \,e^{g(x)} \varphi(x) \vert^2 dx <+\infty.
\label{acota1}
\ee
Hence, from (\ref{eigen1}) one gets
\be
\alpha_1 \varphi'' + (2\alpha_1 g' + \alpha_2) \varphi' +\{
\alpha_1 [g'' + (g')^2] + \alpha_2 g' + \alpha_3 -E\} \varphi=0.
\label{eigen2}
\ee
Now, we introduce a change of the independent variable $x$, ruled
by a bijection $s$ as follows
\be
x \mapsto y=s(x), \qquad y \mapsto x=s^{-1}(y).
\label{indep1}
\ee
The Jacobian of the transformation is given by $J=s'(x)$. If $J
\neq 0$ at a point $x$, the {\it inverse function theorem\/}
indicates that the map $s$ is 1-1 and onto in some neighborhood of
$x$ (see e.g. \cite{Cho77}, pp 91). In this way, to construct a
well defined bijection $s$ we first ask for the involved Jacobian
$s'$ to be free of zeros. On the other hand, let $f$ be a function
of $x$. Then we have:
\be
f(x) = f(s^{-1}(y)) = [f \circ s^{-1}] (y) \equiv f_* (y).
\label{mapeo1}
\ee
Thus, $f_*$ is the representation of the function $f$ in the
$y$-space. In a similar manner we find that $f$ is the
representation of $f_*$ in the $x$-space: $f= [f_* \circ s]$.
Hereafter, and whenever there be no confusion, we drop the
subindex ``$*$'' from the functions in the $y$-representation. The
straightforward calculation departing from Eq.~(\ref{eigen2})
leads to
\be
\alpha_1 (y')^2 \ddot\varphi + [\alpha_1 y'' +(2\alpha_1 g' +
\alpha_2) y'] \dot\varphi +\{ \alpha_1 [g'' + (g')^2] + \alpha_2
g' + \alpha_3 -E\} \varphi=0,
\label{eigen3}
\ee
where $\dot f \equiv df/dy$. This last equation acquires a simple
form if the coefficients of $\ddot \varphi$ and $\dot \varphi$ are
respectively a constant $c_0^2$ (expressed in appropriate units)
and zero. Thereby one has a system of equations
\be
\alpha_1 (y')^2 =c_0^2, \qquad \alpha_1 y'' +(2\alpha_1 g' +
\alpha_2) y' =0,
\label{system}
\ee
the solution of which defines the form of $g$ and $y$ in terms of
the mass position-dependence:
\be
g(x)= \ln \left[ \frac{m(x)}{m_0} \right]^{1/4}, \qquad y= \int
e^{2g(x)} dx +y_0.
\label{gy}
\ee
Here $m_0$ and $y_0$ are integration constants (we shall take, for
simplicity, $y_0=0$) and the constant $c_0$, introduced in
(\ref{system}), has been written as $c_0 = i\hbar/\sqrt{2m_0}$.
These last results in (\ref{eigen3}) reduce the initial eigenvalue
equation (\ref{eigen1}) to the following one:
\be
H_{\rm eff}^{(a)} \varphi (y) :=\left[-\left(
\frac{\hbar^2}{2m_0}\right) \frac{d^2}{dy^2} + V_{\rm
eff}^{(a)}(y)\right] \varphi(y) =E \varphi(y)
\label{eigen4}
\ee
where the function
\be
V_{\rm eff}^{(a)} := V- \left( \frac{\hbar^2}{2m^3} \right) \left[
\left(\frac{1}{4} +a \right) mm'' - \left\{ \frac{7}{16} +a(2+a)
\right\} (m')^2 \right]
\label{effec}
\ee
plays the role of an ``effective potential'' and depends on the
explicit expressions for the mass $m$ and the initial potential
$V$, both of them in the $y$-representation. In particular, if the
mass function $m$ is a constant then we have $V_{\rm eff}^{(a)} =
V$. In general, $m(x)$ could lead to a very complex function
$V_{\rm eff}^{(a)} (y)$ for which the new equation (\ref{eigen4})
is as complicated to solve as the initial one (\ref{eigen1}).
Hence, at this stage, the main simplification is the avoiding of
undesirable mass factors in the derivative term so that techniques
to solve the conventional Schr\"odinger equation can be applied.

Given a solution $\varphi_*$ of (\ref{eigen4}), according with
(\ref{map1}), (\ref{indep1}) and (\ref{gy}), the function $\psi$
is
\be
\psi(x) = J^{1/2}\, [\varphi_* \circ s](x), \quad J =\left[
\frac{m(x)}{m_0} \right]^{1/2}.
\label{map2}
\ee
Therefore we have
\be
\int_{{\cal D}^{(a)}} \vert \psi (x) \vert^2 dx = \int_{{\cal
D}^{(a)}} J\;\! \vert \varphi_* (s(x)) \vert^2  dx \quad
\longleftrightarrow \quad \int_{{\cal D}^{(a)}_{\rm eff}} \vert
\varphi_*(y) \vert^2 dy,
\label{acota2}
\ee
with ${\cal D}_{\rm eff}^{(a)} \equiv Dom (H_{\rm eff}^{(a)})$.
That is, by getting the square-integrable eigenfunctions of
$H^{(a)}_{\rm eff}$ one is able to obtain the square-integrable
eigenfunctions of $H_a$ and viceversa. Moreover, from
(\ref{eigen1}) and (\ref{eigen4}) we notice that $\varphi_*$ and
$\psi$ share the same eigenvalue $E$. Thus, $H_a$ and
$H^{(a)}_{\rm eff}$ are isospectral operators; we write ${\rm
Sp}(H_a) ={\rm Sp} (H^{(a)}_{\rm eff})$.

Notice that equations (\ref{effec}) and (\ref{map2}) are
consistent with the results reported in \cite{Alh02,Gon02,Koc03}.
With regard to our approach, there is yet a couple of special
cases leading to further simplifications. Namely, one can get
$V^{(a)}_{\rm eff}(y) = V(y)$ by selecting the appropriate
function $m(x)$ or by properly fixing the value of $a$, as we are
going to show.

\subsection{Mass-dependent null terms (MDNT)}

Let us look for a mass function $m$ such that $V^{(a)}_{\rm eff}
-V=0$ in Eq.~(\ref{effec}). Thus, we should solve the non-linear,
second order differential equation:
\be
c_1 m m'' + c_2 (m')^2=0, \qquad 2c_1=\frac{1}{16}-a^2-c_2=
\frac12 +2a.
\label{mass1}
\ee
A brief examination yields
\be
m(x;a)=m_0 (x_0 + \lambda x)^{-4/(3+4a)}, \qquad a \neq -3/4
\label{msing}
\ee
as the simplest solution with $x_0$ and $\lambda$ constants to be
fixed. We have to distinguish between two general cases:
\begin{enumerate}
\item[I)]
If $a < -3/4$ then $m(x;a)$ has a zero at $x=t_0 \equiv
-x_0/\lambda$
\item[II)]
If $a > -3/4$ then $m(x;a)$ is singular at $x=t_0$
\end{enumerate}
The first case will be omitted to avoid ill defined operators
$H_a$ and unappropriate mappings $s$ as well. Indeed, if $a<-3/4$,
the kinetic term $K_a$ in Eq. (\ref{hamil1}) diverges and the
Jacobian $J$ in (\ref{map2}) is zero at $x=t_0$. On the other
hand, for $a>-3/4$ the integrability of $\sqrt {m(x;a)}$ in Eq.
(\ref{gy}) depends on the value of $a$. In particular, if $a=a_0
\equiv -1/4$ then the mapping $x \mapsto y_{(0)}$ is ruled by the
function
\be
y_{(0)}= s_{(0)} (x)= \frac{\ln (x_0 + \lambda x)}{\lambda},
\qquad x \geq t_0,
\label{ymsin14}
\ee
with
\be
m_{(0)}(x) \equiv m(x;a_0) = \frac{m_0}{(x_0 + \lambda x)^2}.
\label{m14}
\ee
This last expression of $m(x)$ is connected with the revival
wave-packets in a position-dependent mass infinite well
\cite{Sch06}. Here, the Jacobian reads $J_{(0)} = 1/(x_0+\lambda
x)$, so that the bijection $s_{(0)}$ is well defined for all $x
\geq t_0$ and arbitrary real values of $x_0$ and $\lambda$. Then,
in general ${\cal D}^{(a_0)} \subseteq [t_0,+\infty)$ and ${\cal
D}^{(a_0)}_{\rm eff} \subseteq \mathbb R$. The explicit form of
the domains of definition ${\cal D}^{(a_0)}$ and ${\cal
D}^{(a_0)}_{\rm eff}$ depend on $V$ while the inverse function
reads
\be
x= s_{(0)}^{-1}(y) = \frac{e^{\lambda y_{(0)}}-x_0}{\lambda}.
\label{inverse14}
\ee

On the other hand, if $a \neq a_0$ the new variable is given by
\be
y= s(x;a)= \frac{(x_0 + \lambda x)^{\eta}}{\lambda \eta}, \qquad
\eta = \left( \frac{1+4a}{3+4a} \right).
\label{ymsin}
\ee
The appropriate mapping $s(x;a)$ is fixed by looking for the
values of $a$ such that either $\eta = 2n+1$ or $\eta^{-1}= 2n+1$,
$n=1,2,\ldots$ In the former case we arrive at the discrete set
integrated by the points $-\frac{3n+1}{4n} = -1, -\frac{7}{8},
-\frac{5}{6}, \cdots < -\frac{3}{4}$. However, each one of these
possible values of $a$ belongs to the case (I) discussed above and
must be omitted. Now, if $\eta^{-1} = 2n+1$ we obtain the points
$\frac{1-n}{4n} =0, -\frac{1}{8}, -\frac{1}{6}, \cdots >-
\frac{1}{4}$, which belong to the case (II) we are interested on.
Henceforth, the mapping $x \mapsto y_{(n)}$ is established from
Eq. (\ref{ymsin}) as follows
\be
y_{(n)} = s(x;a_n) \equiv s_{(n)} (x) = \left(
\frac{2n+1}{\lambda} \right) (x_0 + \lambda x)^{1/(2n+1)}, \quad
a_n \neq a_0,
\label{ysmin2}
\ee
while the corresponding inverse transformation is ruled by
\be
x=  s_{(n)}^{-1}(y_{(n)})= \frac{1}{\lambda} \left[ \left(
\frac{\lambda y_{(n)}}{2n+1} \right)^{2n+1} -x_0 \right], \quad
a_n \neq a_0.
\label{inverse}
\ee
The expression for the mass-function (\ref{msing}) in terms of
$a_n \neq a_0$ reduces to
\be
m_{(n)}(x) \equiv m(x;a_n)= \frac{m_0}{(x_0+\lambda
x)^{4n/(2n+1)}}, \quad n\in \mathbb{N}.
\label{msing2}
\ee
Remark that $J_{(n)} = (x_0 + \lambda x)^{-2n/(2n+1)}$. Hence
$J_{(n)} \neq 0$ for all $x \in \mathbb{R}$ and arbitrary real
values of $x_0$ and $\lambda$. As a consequence ${\cal D}^{(a_n)}
\subseteq \mathbb R$ and ${\cal D}^{(a_n)}_{\rm eff} \subseteq
\mathbb R$. To embrace $a_{n>0}$ and $a_0$ into the same notation
let us introduce the set
\be
{\cal A} = \left\{ a_0 =-1/4, a_n = \frac{1-n}{4n} \right\},
\quad n \in \mathbb{N}.
\label{set3}
\ee
Then, if $a \in {\cal A}$ the position dependent mass operator
$H_{a_n}$ is mapped to a conventional Hamiltonian $H^{(a_n)}$ in
the $y_{(n)}$-representation and viceversa (see Table~\ref{tab1}):
\be
H_{a_n} \, \leftrightarrow \, H_{\rm eff}^{(a_n)} \equiv H^{(a_n)}
= -\left( \frac{\hbar^2}{2m_0} \right) \frac{d^2}{dy_{(n)}^2} +
V(y_{(n)}),
\label{free}
\ee
with $\textrm{Sp} (H_{a_n}) = \textrm{Sp} (H^{(a_n)})$. We shall
take full advantage of this last property in the next sections.

\subsection{Mass-independent null terms (MINT)}

A simple inspection to equation (\ref{effec}) shows that $V_{\rm
eff}^{(-1/4)} (y) = V(y)$, no matter the explicit form of the mass
function $m(x)$ -assuming this last is well defined-. That is, by
fixing $a=-1/4$ we get:
\be
H_{-1/4} \leftrightarrow H_{\rm eff}^{(-1/4)} \equiv H = -\left(
\frac{\hbar^2}{2m_0} \right) \frac{d^2}{dy^2} + V(y).
\label{free2}
\ee
In particular, if $m= m_{(0)}$ then $H$ in (\ref{free2}) is the
same as $H^{(a_0)}$ with ${\cal D}^{(-1/4)} \equiv \textrm{Dom}
(H_{-1/4})={\cal D}^{(a_0)}$ and $\textrm{Dom}(H) = {\cal
D}^{(a_0)}_{\rm eff}$. A similar situation occurs if $m=m_{(n)}$
(see Table~\ref{tab1}).

Besides the mass-functions derived in the previous section, a
regular expression for $m$ has been recently introduced in
\cite{Cru07} (see also \cite{Car04}). This is given by the
function $m_R$:
\be
m_R(x) = \frac{m_0}{1+(\lambda x)^2}, \qquad \lambda \in
\mathbb{R}
\label{mass}
\ee
with
\be
s_R(x) = \frac{{\rm arcsinh}(\lambda x)}{\lambda}, \qquad
s_R^{-1}(y) = \frac{\sinh (\lambda y)}{\lambda}.
\label{ese}
\ee
The corresponding Jacobian $J=1/\sqrt{1 +(\lambda x)^2}$ is
nonzero for all $x\in \mathbb{R}$ and arbitrary values of
$\lambda$. Hence ${\cal D}^{(-1/4)} \subseteq \mathbb R$ and
${\cal D}^{(-1/4)}_{\rm eff} \subseteq \mathbb R$. The main
aspects of these last results are summarized in Table~\ref{tab1}.
As a final remark, the mass (\ref{mass}) appeared in the
construction of the Wigner functions connected with a class of
position-dependent oscillators \cite{Des08}. Other interesting
mass-functions are
\be
m_w(x) = \left( \frac{w+x^2}{1+x^2} \right)^2, \qquad m_c(x) =
cx^2.
\label{masses}
\ee
They have been already studied in \cite{Pla99} and recently in
e.g. \cite{Des00}. Notice that $J_c (x=0) =0$ and $J_w \neq 0$
$\forall x\in \mathbb R$ and $w>0$. In the next sections we shall
study specific forms of the potentials $V(x)$ and $V_*(y)$ which
represent oscillator-like interactions for a position-dependent
mass quantum system.

\vskip2ex
\begin{table}
\centering {\footnotesize
\begin{tabular}{||l|c|c||}
\hline
\hline
$a_0 = -\frac{1}{4}$ & $H_{a_0}=\frac12 \, m_{(0)}^{-1/4} P\,
m_{(0)}^{-1/2} P\, m_{(0)}^{-1/4} + V(x)$ & $H^{(a_0)}=-\left(
\frac{\hbar^2}{2m_0} \right)
\frac{d^2}{dy_{(0)}^2} + V(y_{(0)})$\\
$m_{(0)}(x)= \frac{m_0}{(x_0 + \lambda x)^2}$ &&\\
$\lambda y_{(0)} = \ln(x_0+\lambda x)$ & ${\cal D}^{(a_0)}
\subseteq [t_0, +\infty)$ & ${\cal D}^{(a_0)}_{\rm eff}
\subseteq \mathbb R$\\
\hline
$a_1 = 0$ & $H_{a_1}=\frac12 \, P \, m_{(1)}^{-1} P + V(x)$ &
$H^{(a_1)}=-\left( \frac{\hbar^2}{2m_0} \right)
\frac{d^2}{dy_{(1)}^2} + V(y_{(1)})$\\
$m_{(1)}(x) = \frac{m_0}{(x_0+ \lambda x)^{4/3}}$ &&\\
$\lambda y_{(1)} = 3 (x_0 + \lambda x)^{1/3}$ & ${\cal D}^{(a_1)}
\subseteq \mathbb R$ & ${\cal D}^{(a_1)}_{\rm eff}
\subseteq \mathbb R$\\
\hline
$a_n = \frac{(1-n)}{4n}, n\in \mathbb{N}$ & $H_{a_n}=\frac12 \,
m_{(n)}^{a_n} P\, m_{(n)}^{-1-2a_n} P\, m_{(n)}^{a_n} + V(x)$  &
$H^{(a_n)}=-\left(
\frac{\hbar^2}{2m_0} \right) \frac{d^2}{dy_{(n)}^2} + V(y_{(n)})$\\
$m_{(n)}(x)$ -see Eq.~(\ref{msing2})- &&\\
$y_{(n)}$ -see Eq.~(\ref{ysmin2})- & ${\cal D}^{(a_n)} \subseteq
\mathbb R$ & ${\cal D}^{(a_n)}_{\rm eff}
\subseteq \mathbb R$\\
\hline
$a= -\frac{1}{4}$ & $H_{-1/4}=\frac12 \, m^{-1/4} P\, m^{-1/2} P\,
m^{-1/4} + V(x)$ & $H=-\left(
\frac{\hbar^2}{2m_0} \right) \frac{d^2}{dy^2} + V(y)$\\
$m(x) >0$ &&\\
$y = \int (m/m_0)^{1/2} dx$ & ${\cal D}^{(-1/4)} \subseteq \mathbb
R$ & ${\cal D}^{(-1/4)}_{\rm eff} \subseteq \mathbb R$\\
\hline
\hline
\end{tabular}
\caption{\footnotesize Special mass functions $m(x)$ and orderings of the
kinetic term $K_a(x)$ leading to the Hamiltonians $H_a$ and
$H^{(a)}_{\rm eff}$, with $V^{(a)}_{\rm eff} (y)= V(y)$ and ${\rm
Sp}(H_a)={\rm Sp}(H^{(a)}_{\rm eff})$. In all cases the definite
domain is fixed by $V$.}
\label{tab1}
}
\end{table}

\section{Two kinds of position-dependent mass oscillators}
\label{sec3}

We are going to work with the eigenvalue equation (\ref{eigen4})
such that $V_{\rm eff}^{(a)} = V$ by either the MDNT or the MINT
cases described in the previous sections. Although our approach
holds for any well defined potential $V$, we shall focus on the
linear harmonic oscillator in two general situations:

\begin{itemize}
\item[i)] Departing from a given interaction $V(x)$ and a mass
function $m(x)$ we arrive at the conventional linear harmonic
oscillator problem in the $y$-representation. That is, the new
potential reads $V_*(y) =\frac{m_0 \omega_0^2}{2}\, y^2$, with
$\omega_0$ the natural frequency of oscillation. Since $V(x)$ and
$V_*(y)$ are isospectral they share the eigenvalues defined by
$E_n= \hbar \omega_0 (n+1/2)$, $n=0,1,2,\ldots$ We shall refer to
these potentials as {\em oscillators of the first kind}.

\item[ii)] Departing from the linear harmonic oscillator
interaction $V(x) =\frac{m_0 \omega_0^2}{2}\, x^2$ and a mass
function $m(x)$ we arrive at the eigenvalue equation connected
with the new potential $V_*(y)$. Since $V(x)$ and $V_*(y)$ are
isospectral we solve the (conventional) Schr\"odinger equation in
the $y$-representation to construct the solutions of the initial
oscillator-like, position-dependent mass problem. We shall refer
to these potentials as {\em oscillators of the second kind}.

\end{itemize}

\subsection{Oscillators of the first kind}
\label{sec31}

Let us take $V_*(y) =\frac{m_0 \omega_0^2}{2}\, y^2$ as the
$y$-representation of the initial potential $V(x)$. Then ${\cal
D}^{(a)}_{\rm eff} = \mathbb R$ and all the mappings MDNT and MINT
can be applied (see Table~\ref{tab1}). It is convenient to
introduce a dimensionless notation as follows:
\be
\left[ -\frac12 \frac{d^2}{d\textrm{y}^2} + \frac{\textrm{y}^2}{2}
-\textrm{E} \right] \varphi(\textrm{y})=0, \qquad  \textrm{y} = y
\left( \frac{\hbar}{m_0 \omega_0} \right)^{-1/2} \equiv y \alpha,
\quad \textrm{E}= \frac{E}{\hbar \omega_0}
\label{free1}
\ee
Then, the solutions read
\be
\varphi_n(\textrm{y}) = \frac{H_n(\textrm{y})
e^{-\textrm{y}^2/2}}{\sqrt{2^n \pi^{1/2}k!}}, \quad
H_n(\textrm{y}) = (-1)^n e^{\textrm{y}^2/2}
\frac{d^n}{d\textrm{y}^n} e^{-\textrm{y}^2/2}, \quad
\textrm{E}_n=n+ \frac12.
\label{dim1}
\ee
Next, we are going to solve the initial position-dependent mass
problem in terms of these results.

\subsubsection{MDNT case}

Let $m(x;a)$ be the mass function with $a \in {\cal A}$, that is
$m=m_{(n)}$, $n=0,1,2,\ldots$ From equations (\ref{ymsin14}) and
(\ref{ysmin2}) we know that the initial potential reads
\be
V_{(n)}(x)=[V_* \circ s_{(n)}](x) = \frac{m_0 \omega_0^2}{2
\lambda^2} \left\{
\begin{array}{cl}
(2n+1)^2 (x_0 + \lambda x)^{\frac{2}{(2n+1)}}, & n \in \mathbb{N},
\, x \in \mathbb{R}\\[2ex]
\ln^2 (x_0 + \lambda x), & n=0, \, x\in [t_0,+\infty)
\end{array}
\right.
\label{resing1}
\ee
A dimensional analysis shows that $\lambda = \lambda_0 \alpha$,
with $\lambda_0$ a constant ($\lambda_0=1$ for simplicity). The
behavior of potential (\ref{resing1}) for $n=1$ and $n=2$ is
contrasted with the well known curve of the harmonic oscillator
potential in Fig.~\ref{figpotq}a; the case $n=0$ is depicted in
Fig.~\ref{figpotq}b. In both cases, as we have previously noted,
the involved spectrum is given by $E_k=\hbar \omega_0 (k+1/2)$,
$k=0,1,2,\ldots$, while their eigenfunctions respectively read
\be
\psi_k(x) = \frac{H_k[(2n+1)(x_0+\alpha
x)^{\frac{1}{(2n+1)}}]}{(x_0+\alpha x)^{\frac{n}{(2n+1)}}\sqrt{2^k
\pi^{1/2}k!}} \exp \left[ -\frac{(2n+1)^2}{2} (x_0+\alpha
x)^{\frac{2}{(2n+1)}}\right], n\in\mathbb{N},
\label{eigv1}
\ee
and
\be
\psi_k(x) = \frac{H_k[\ln (x_0 + \alpha x)]}{\sqrt{(x_0 + \alpha
x)\, 2^k \pi^{1/2}k!}}\, e^{-\frac12 \ln^2(x_0 + \alpha x)}, \quad
n=0,
\label{eigv2}
\ee
with $x$ running in the domains indicated in (\ref{resing1}).
Thus, the energy spectrum of a position-dependent mass quantum
system which is subject to either the action of a confining
odd-root-law potential $V_{(n)}(x) \propto (\alpha x)^{2/(2n+1)}$,
$n \in \mathbb{N}$, or to a square-logarithmic interaction
$V_{(0)}(x) \propto \ln^2(\alpha x)$, is ruled by the quantization
of the conventional harmonic oscillator energy if the
mass-function is respectively taken as $m_{(n)}$ or $m_{(0)}$.

\subsubsection{MINT case}

\begin{figure}[h]
\centering
\includegraphics[width=6.8cm]{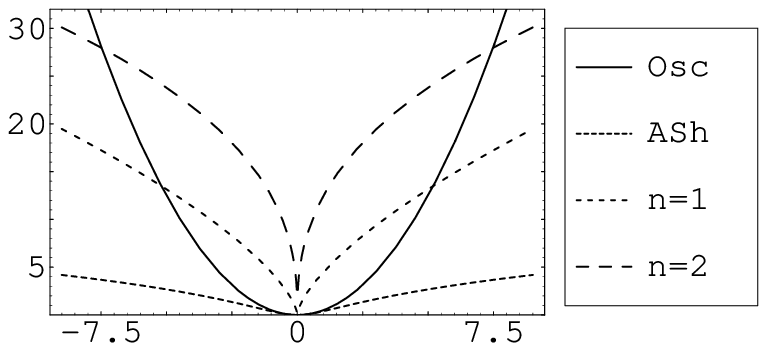} \hskip.5cm
\includegraphics[width=7cm]{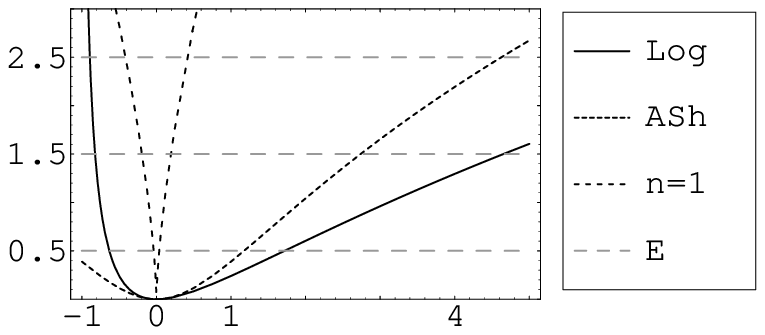}\\
(a) \hskip6cm (b)
\caption{\footnotesize (a) The odd root-law potential (\ref{resing1})
with $n=1$, $n=2$ and $x_0=0$ besides the regular one (ASh)
defined in Eq. (\ref{resorte1}). The harmonic oscillator potential
(Osc) is depicted as a reference. (b) The square-logarithmic
potential (Log) defined in Eq. (\ref{resing1}) with $x_0=1$ and
${\cal D}^{(a_0)} = [-1,+\infty)$. Potentials (Ash) and (n=1) as
well as the first three energy levels (E) are also depicted.
Vertical and horizontal axis are respectively in $\hbar \omega_0$
and dimensionless units.}
\label{figpotq}
\end{figure}

Let $m_R(x)$ be the mass function with $\lambda =\alpha$. Then,
the spectrum of the potential
\be
V(x) = [V_* \circ s](x) = \frac{\hbar \omega_0}{2} \,
\textrm{arcsinh}^2(\alpha x)
\label{resorte1}
\ee
is given by $E_k=\hbar \omega_0 (k+1/2)$, $k=0,1,2,\ldots$, and
the involved eigenfunctions read
\be
\psi_k(x) = \left[ \frac{m(x)}{2^{2k} \pi (k!)^2 \, m_0}
\right]^{1/4} H_k [\textrm{arcsinh}(\alpha x)] \, e^{-\frac{1}{2}
\,\, \textrm{arcsinh}^2(\alpha x)}, \quad k=0,1,2,\ldots
\label{eigenreg1}
\ee
In Fig.~\ref{figpotq} the global behavior of potential
(\ref{resorte1}) is shown in contrast with the curve of the
harmonic oscillator one.

As we can see, one is  able to identify the kind of interaction
$V(x)$ which has to be applied to a quantum system of
position-dependent mass $m(x)$ in order supply it with a specific,
well known, spectrum $\textrm{Sp}(H^{(a)})$. For instance, if
$\textrm{Sp}(H^{(a)}) = \{\hbar \omega_0
(n+1/2)\}_{n=0}^{+\infty}$, we have shown that the system has to
be subject to potentials behaving as a confining odd-root-law,
$\ln^2$ or $\sinh^2$, whenever the mass-function is respectively
defined by (\ref{m14}), (\ref{msing2}) or (\ref{mass}). A more
deeper insight is necessary if one is interested on a
position-dependent mass $m(x)$, subject to a particular
interaction $V(x)$, rather than in the recovering of a given
spectrum. That is, what the sort of the spectrum is expected by
applying an oscillator-like interaction to a quantum system of
mass-function $m(x)$? The problem is going to be faced in the next
section.

\subsection{Oscillators of the second kind}

In this section we analyze the effects on the energy spectrum
produced by a position dependence of the mass. In other words, how
different is the spectrum of a system of mass $m(x)$ from that of
a particle of mass $m_0$ when both of them are under the action of
the same potential $V(x)$? As before, we shall focus on the
simplest case of the linear harmonic oscillator interaction.

Let $V_{\rm osc}(x) =\frac{m_0 \omega_0^2}{2} \, x^2$ be the
initial potential. Notice that $Dom (V_{\rm osc}) = \mathbb R$
requires ${\cal D}^{(a)} = \mathbb R$. However ${\cal D}^{(a_0)}
\subseteq [t_0, +\infty)$, so that $a$ must be different from
$a_0$ (see Table~\ref{tab1}). The case $a=a_0$ will be analyzed in
Section~\ref{seceqz}.

\subsubsection{MDNT case}

\begin{figure}[h]
\centering
\includegraphics[width=6cm]{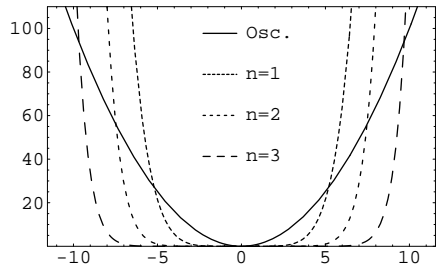}
\caption{\footnotesize Three members of the family of power-law potentials
(\ref{resing}). The conventional harmonic oscillator potential
(Osc.) is recovered for the forbidden value $n=0$ and the family
goes to the free particle case for $n \rightarrow +\infty$.}
\label{figpot}
\end{figure}

Let $m_{(n)}(x)$ be the mass function with $n\in \mathbb{N}$ and
$\lambda = \alpha$. The initial potential $V_{\rm osc}(x)$ behaves
as an even-power-law function in the $y_{(n)}$-space (see
Fig.~\ref{figpot}):
\be
V_*(y;n) = [V_{\rm osc} \circ s_{(n)}^{-1}](y) = \frac{\hbar
\omega_0}{2} \left[ \left( \frac{\alpha y}{2n+1} \right)^{2n+1} -
x_0\right]^2, \quad n\in \mathbb{N}
\label{resing}
\ee
where the label ``$(n)$'' has been dropped from the $y$-coordinate
for simplicity. Hereafter we shall take $x_0=0$. Notice that
$V_*(y;n) \rightarrow 0$ as $n \rightarrow +\infty$ and $V_*(y;n)
\rightarrow \hbar \omega_0 (\alpha y)^2/2$ as $n \rightarrow 0$.
Thus, the family of potentials (\ref{resing}) is delimited by the
free particle and the harmonic oscillator potentials (remember
that $n=0$ and $n\rightarrow +\infty$ are forbidden in
Eq.~\ref{ysmin2}). Such a behavior is shown in
Figure~\ref{figpot}.

Let us emphasize that, although this kind of potentials is not
analytically solvable, they have deserved special attention in
pedagogical as well as in research papers over the years. For
instance, their WKB energy levels have been shown to depend on the
power of the potential \cite{Suk73} and the involved scale
invariance has been studied in terms of the Lie method
\cite{Car93}. The polarizability of a particle in a power-law
potential due to the presence of a constant force and the wave
packet revivals in such potentials, on the other hand, have been
exhaustively studied in \cite{Rob98} and \cite{Rob00}
respectively.

\begin{figure}[h]
\centering
\includegraphics[width=6cm]{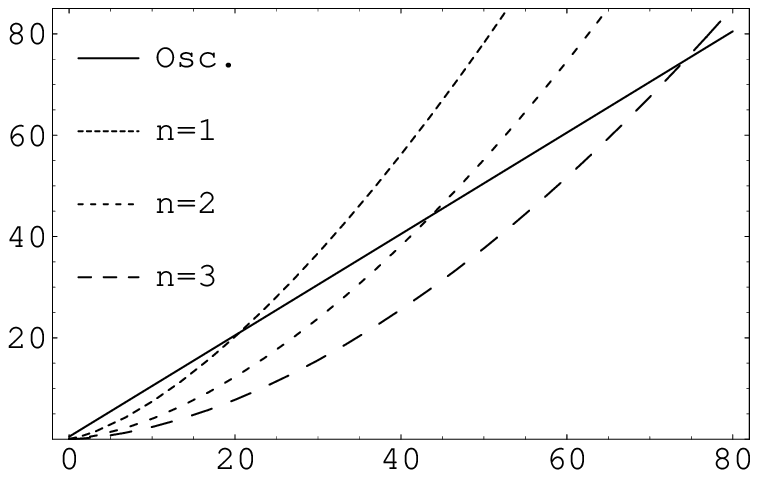} \hskip1cm
\includegraphics[width=6cm]{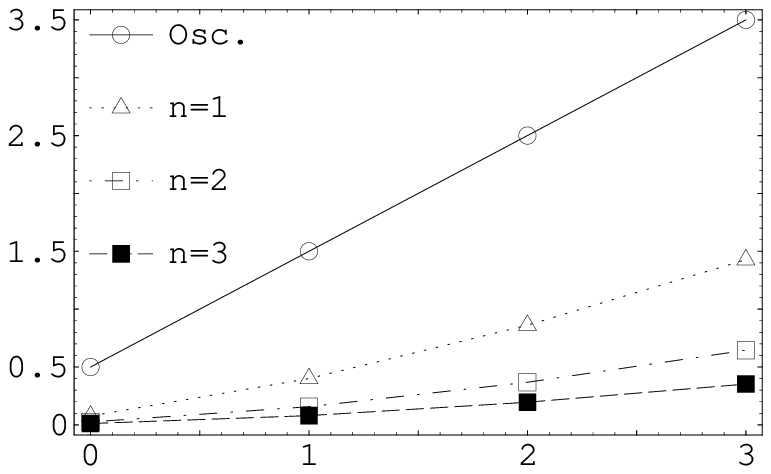}
\caption{\footnotesize (Left) The spectrum curves $E_k(n)$ of the
power-law potential (\ref{resing}) for $n=1,2,3$, besides the
spectrum curve of the harmonic oscillator (Osc.); all of them are
depicted in $\hbar \omega_0$ units as a function of $k$. Notice
the points in which $E_k(n) = E_k (0)$. (Right) Details of the
first four energy levels.}
\label{figspec}
\end{figure}

In general, the roots of $V_*(y;n)=V_{\rm osc}(y)$ define a region
$(-y_c,y_c) \subset {\cal D}^{(a_n)}_{\rm eff}$ in which the
potential $V_*$ grows up slower than $V_{\rm osc}$. The geometry
of these last curves in $(-y_c,y_c)$ suggest the spectrum of $V_*$
will be integrated by energy levels which are below the
corresponding oscillator energies. The behavior of the curves in
the complementary region ${\cal D}^{(a_n)}_{\rm
eff}\setminus(-y_c,y_c)$ is such that the energy levels are
expected to be above the oscillator ones. To verify our statement
let us calculate the eigenvalues of $V_*(y;n)$ by means of the
energy quantization condition of the WKB method:
\be
\int_{-y_0}^{+y_0} \sqrt{2m_0[E(n) -V(y;n)]} \, dy=\pi \hbar
(k+1/2), \quad k=0,1,2,\ldots
\label{wkb1}
\ee
with $\pm y_0 =\pm (\frac{2n+1}{\alpha}) (2
\textrm{E}(n))^{1/(4n+2)}$ the classical (symmetric) turning
points and $E(n)$ the energy connected with the potential
$V_*(y;n)$ for a given $n \in \mathbb N$. The change of variable
$y =y_0 z$ reduces the integral equation (\ref{wkb1}) to (compare
with \cite{Suk73} and \cite{Car93}):
\be
E_k(n)= \frac{\hbar \omega_0}{2} \left[ \frac{\pi}{j_n}
\frac{(k+1/2)}{(2n+1)} \right]^{\frac{2n+1}{n+1}}
\label{wkb2}
\ee
where the constant
\be
j_n = \int_{-1}^1 \sqrt{1-z^{4n+2}}dz = \frac{\sqrt{\pi} \,
\Gamma\left( \frac{1}{4n+2} \right)}{2(n+1) \Gamma \left(
\frac{n+1}{2n+1} \right)}
\label{wkb3}
\ee
is such that $j_n \rightarrow 2$ as $n \rightarrow +\infty$ and
$j_0=\pi/2$. Figure~\ref{figspec} shows the spectrum curves of
three members of the family (\ref{resing}) compared with the
energy spectrum curve of the harmonic oscillator. Notice that the
energy levels become closer to each other as the label $n$
increases (free particle case). That is, if $n>>1$ then $E_k
(n)\propto [(k+1/2)/n]^2$. On the other hand, for the forbidden
value $n=0$ we have the oscillator spectrum $E_k(0) =\hbar
\omega_0 (k+1/2)$, as it was expected. The corresponding set of
eigenfunctions, in turn, can be numerically constructed or
analyzed by using improved versions of the WKB method like that
discussed in \cite{Fri96}.

Now, let us take one of the allowed values of $n$. The root of
equation $E_k(n) = E_k(0)$ is given by
\be
k_c (n)= \frac12 \left\{ \left[ \frac{(2n+1)
\Gamma(\frac{1}{4n+2})}{\sqrt{\pi} (n+1)\Gamma(\frac{n+1}{2n+1})}
\right]^{\frac{2n+1}{n}} -1\right\}.
\label{root}
\ee
The ceiling function $\lceil k_c \rceil$ of $k_c(n)$ defines a
subset of $\textrm{Sp}(V_*(y;n))=\{ E_k(n) \}_{k=0}^{+\infty}$ for
which $E_k(n)<E_k(0)$ $\forall k < \lceil k_c \rceil$. The larger
the value of $n$ the bigger the set of eigenvalues $E_k (n)$
delimited by $E_k (0)$. The complementary set is then such that
$E_k(n)>E_k(0)$ $\forall k \geq \lceil k_c \rceil$; details are
shown in Figure~\ref{figspec}.

In conclusion, the oscillator of the second kind defined by the
pair $(V_{\rm osc}, m_{(n)})$, $n\in \mathbb{N}$, shares its
spectrum with a particle of mass $m_0$ subject to an
even-power-law potential of the form $V_{*}(y_{(n)};n) \propto
[y_{(n)}/(2n+1)]^{4n+2}$. When contrasted with a conventional
oscillator of mass $m_0$, the energy spectrum of the pair $(V_{\rm
osc}, m_{(n)})$ is a {\em distorted} version of the oscillator
one. The shape and amount of the distortion are respectively
dictated by Eq. (\ref{wkb2}) and $k_c(n)$, as this last was
defined in (\ref{root}). That is, the distortion is stronger for
larger values of $\vert \lceil k_c\rceil -k\vert$ in equation
(\ref{wkb2}).

\subsubsection{MINT case}
\label{322}

\begin{figure}
\centering
  \includegraphics[width=6cm]{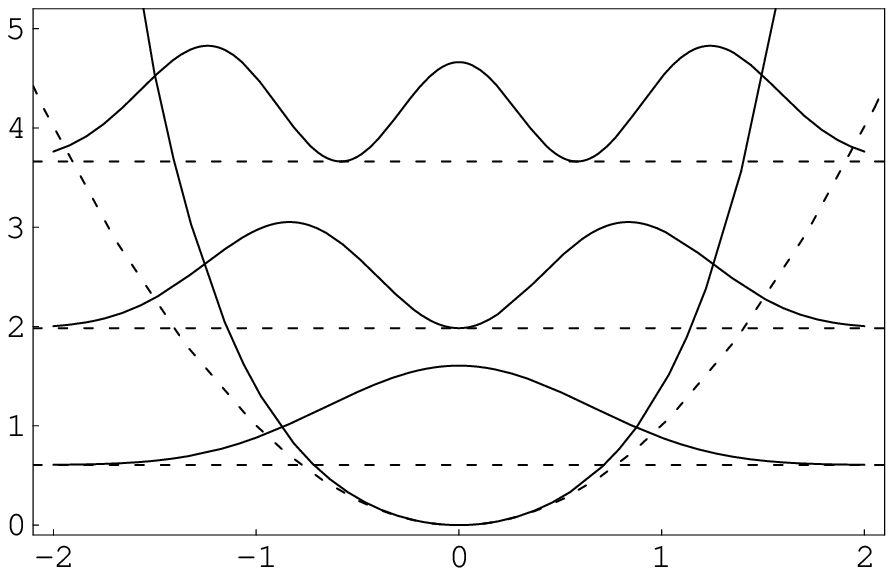}
  \hskip1cm
\includegraphics[width=6cm]{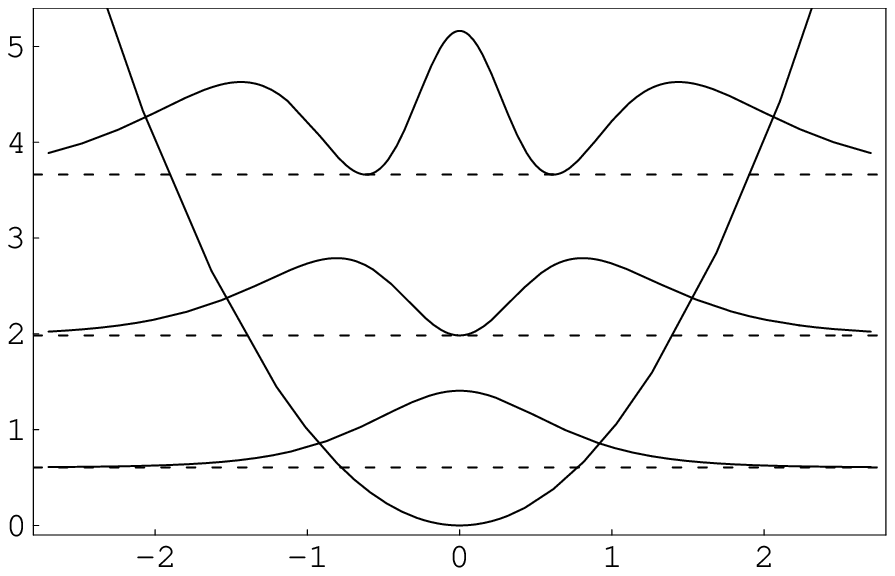}
\caption{\footnotesize (Left) The potential $\sinh^2(\textrm{y})$ and the probability
densities of its three first wavefunctions together with the
corresponding energy levels. The oscillator potential curve
(dashed) is included as a reference. (Right) Probability densities
of the first three wavefunctions of a second kind,
position-dependent mass oscillator which shares its spectrum with
the $\sinh^2(\textrm{y})$ potential. In both cases the vertical
and horizontal axis are respectively in $\hbar \omega_0$ and
dimensionless units.}
\label{sinhfig}
\end{figure}

Let $m_R(x)$ be the mass function with $\lambda =\alpha$. The
potential in the $y$-representation reads
\be
V_*(y) = [V_{\rm osc} \circ s_R^{-1}](y) =\frac{\hbar \omega_0}{2}
\, \textrm{sinh}^2(\alpha y).
\label{resorte2}
\ee
Here, the (dimensionless) Schr\"odinger equation to solve is
\be
\left[ -\frac{d^2}{d\textrm{y}^2} + \textrm{sinh}^2 \textrm{y}
\right] \varphi = 2\textrm{E} \varphi.
\label{schro3}
\ee
As in the previous case, the energy quantization condition
(\ref{wkb1}) gives an accurate approximation to the eigenvalues
$\textrm{E}$ of the energy. With the classical turning points $\pm
y_0 = \pm \mathrm{arcsinh} (\sqrt{2 {\rm E}})/\alpha$, one arrives
at the following transcendental equation:
\be
\sqrt{2\mathrm{E}}\; F_E \left(i\, \mathrm{arcsinh}\,
\sqrt{2\mathrm{E}}\,\left\vert -\frac{1}{2\mathrm{E}} \right.
\right) = i\, \frac{\pi}{2} (k+1/2)
\label{sinesp}
\ee
where
\[
F_E (\varphi \mid m ) = \int_0^{\varphi} ( 1-m \sin^2\theta)^{1/2}
d\theta
\]
is the {\it Elliptic Integral of the Second Kind\/} (see e.g.
\cite{Abr72}). The roots $\textrm{E}_k$ of (\ref{sinesp}) can be
evaluated numerically by using conventional algorithms. In
Table~\ref{table2} we show some of the first values of $E_k$
compared with those obtained from a direct, numerical integration
of the Schr\"odinger equation (\ref{schro3}). The corresponding
probability densities $\vert \varphi_*(\textrm{y}) \vert^2$ are
plotted in Figure~\ref{sinhfig}, contrasted with their partners
$\vert \psi (x) \vert^2$ in the $x$-representation.

\begin{table}
\centering
\begin{tabular}{|c|r|r|}
  \hline
   & \multicolumn{2}{|c|}{$E_k$ in $\hbar w_0$ units} \\
   \cline{2-3}
  k & \multicolumn{1}{|c|}{WKB} & \multicolumn{1}{|c|}{Schr\"odinger}\\
  \hline\hline
  0  & 0.55644 & 0.60571 \\
  1  & 1.94482 & 1.98368 \\
  2  & 3.62813 & 3.66250 \\
  3  & 5.56179 & 5.59365 \\
  4  & 7.71941 & 7.74948 \\
  5  & 10.08292 & 10.11165 \\
  6  & 12.63890 & 12.66657 \\
  7  & 15.37683 & 15.40365 \\
  8  & 18.28821 & 18.31431 \\
  9  & 21.36592 & 21.39141 \\
  \hline\hline
\end{tabular}
\caption{\footnotesize The first ten energy levels of the potential
$\sinh^2(\textrm{y})$ calculated numerically from the WKB
transcendental equation (\ref{sinesp}) and directly from the
Schr\"odinger equation (\ref{schro3}).}
\label{table2}
\end{table}

In this case the geometry of the curves $V_*(y)$ and $V_{\rm
osc}(y)$ is in correspondence with the fact that all the energy
eigenvalues of $V_*$ are above the related energy levels of
$V_{\rm osc}$. Indeed, around the origin one has $V_* \gtrsim
V_{\rm osc}$, so that $E_0 \gtrsim 0.5 \hbar \omega_0$, as
expected (see Table~\ref{tab1}). For an arbitrary excited level
$E_k$, the distortion is as strong as fast is the growing up of
$V_*-V_{\rm osc}$. In conclusion, the oscillator of the second
kind $(V_{\rm osc}, m_R)$ shares its spectrum with a particle of
mass $m_0$ subject to the $\sinh^2$ potential (see
Fig.~\ref{sinhfig}). The spectrum, in turn, is a strong distorted
version of the conventional oscillator's one.

\subsubsection{The squeezed oscillator}
\label{seceqz}

\begin{figure}
\centering
\includegraphics[width=6cm]{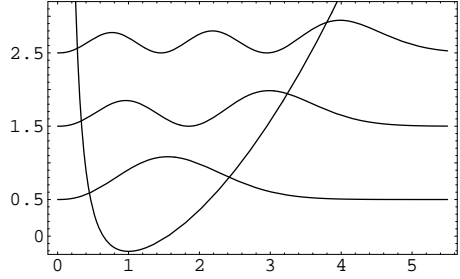}
\caption{\footnotesize The squeezed oscillator (\ref{spec1}) and its first three
probability densities. Vertical and horizontal axis are in $\hbar
\omega_0$ and dimensionless units respectively.}
\label{figsqz}
\end{figure}

Let us consider the potential
\be
V_{\rm sq}(x) = \frac{\hbar \omega_0}{8} \left\{ \left[
\frac{1}{x_0 + \alpha x} -(x_0 +\alpha x) \right]^2 +2(1-\sqrt{2})
\right\}, \qquad x\geq t_0
\label{spec1}
\ee
with $\alpha$ defined in (\ref{free1}), $x_0$ a dimensionless
constant and $\textrm{Dom}(V_{\rm sq}) = [t_0,+\infty)$. This
potential is often refered as the ``singular oscillator'' because
its singularity at $x=t_0$. The conventional expression $V_{\rm
sq}(x)= m_0 \omega_0^2 \left( \frac{x^2}{2} + \frac{g^2}{x^2}
\right)$, with $g$ in units of the square of distance and shifted
by $-\sqrt{2}\hbar \omega_0/4$, is recovered from (\ref{spec1})
with $x_0 + \alpha x =z$ and $\alpha^2 g=\sqrt{2}/4$. Here, we
prefer to call it {\it sqeezed oscillator\/} because its domain of
definition is the result of a L.H.S. `squeezing' of $\mathbb R$ in
terms of $s^{-1}$, as it was established in the previous sections.
If the mass function $m(x)$ is a constant $m_0$ the involved
(dimensionless) Schr\"odinger equation
\be
-\frac{d^2}{dz^2} \varphi + \frac{1}{4} \left[ \left(\frac{1}{z}
-z \right)^2 -2(\sqrt{2}-1) \right] \varphi =2\textrm{E} \varphi,
\qquad z=x_0 +\alpha x
\label{Sspec}
\ee
can be solved in terms of confluent hypergeometric functions by
means of the appropriate transformation (see e.g. \cite{Neg00b}).
Indeed, the mapping $\varphi \rightarrow z^\ell e^{-z^2/4} u(z)$,
$z\mapsto \sqrt{2\chi}$, leads to the following Kummer equation
\be
\chi
\label{Sspec2} \frac{d^2}{d\chi^2} u + \left( \frac{2+\sqrt{2}}{2}
- \chi \right) \frac{d}{d\chi} u - \left( \frac{1}{2} - \textrm{E}
\right) u =0,
\ee
with $\ell=(1+\sqrt{2})/2$. Thereby, the physical solutions for
$E_n=\hbar \omega_0 (n+1/2),\; n=0,1,2,\ldots$, read
\be
\begin{array}{rl}
\varphi_n(z)= & C_n \, z^{\frac{1+\sqrt{2}}{2}} \,
e^{-\frac{z^2}{4}} {}_1F_1 \left(
-n,1+\frac{1}{\sqrt{2}},\frac{z^2}{2} \right)\\[2ex]
= & \left(\frac{n!}{2^{1/\sqrt{2}}
\Gamma(n+1+1/\sqrt{2})}\right)^{1/2} z^{\frac{1+\sqrt{2}}{2}} \,
e^{-\frac{z^2}{4}} L_n^{(1/\sqrt{2})} (z^2/2)
\end{array}
\label{spfun}
\ee
with $L^{(\alpha)}_n(x)$ the {\it Generalized Laguerre
Polynomials\/} \cite{Abr72}. So that, for a constant-mass quantum
system, the one-dimensional potential (\ref{spec1}) shares its
spectrum with the conventional linear harmonic oscillator.
Moreover, it is well known that formulae (\ref{spfun}) can be also
algebraically obtained in terms of the $su(1,1)$ Lie algebra (see
e.g. \cite{Per86}, pp 217). The squeezed oscillator is shown in
Figure~\ref{figsqz}, together with some of the corresponding
probability densities.

If $m(x)$ is not a constant then the energy spectrum of the
quantum system is modified, as we have previously verified.
According to Table~\ref{tab1}, any of the masses (\ref{m14}),
(\ref{msing2}) or (\ref{mass}) allows the mapping to the
$y$-space. First let us consider the case $m=m_{(0)}$. The
potential (\ref{spec1}) is mapped to the following one
\be
V_*(y_{(0)}) = [V_{\rm sq} \circ s_{(0)}^{-1}](y_{(0)}) =
\frac{\hbar \omega_0}{2} \left[\sinh^2(\alpha y_{(0)})
+\frac{1-\sqrt{2}}{2} \right]
\label{spec2}
\ee
which, up to an additive constant, is the same as the potential
reported in Eq. (\ref{resorte2}). Thereby, we have shown that a
quantum system endowed with mass $m_{(0)} (x)$ and acted by the
oscillator-like potential (\ref{spec1}) shares its spectrum with a
particle of mass $m_0$ which is under the action of the potential
$\sinh^2(\textrm{y}_{(0)})$. In comparison with a constant mass
quantum oscillator, we realize that the presence of $m_{(0)} (x)$
distorts the ground energy level of the oscillator-like system
from $0.5$ to $\approx 0.6$ energy units ($\hbar \omega_0$), the
second one from $1.5$ to $\approx 1.9$ and so on. The higher the
level of excitation of the system the stronger the distortion of
the spectrum.

Notice that the system studied in Section~\ref{322} behaves in a
similar manner, so there exists a clear relationship between
position-dependent mass systems: different masses combined with
appropriate interactions give rise to the same spectrum. In this
case, the oscillators of the second kind defined by the pairs
$(V_{\rm osc}, m_R)$ and $(V_{\rm sq}, m_{(0)})$ are isospectral
(see Table~\ref{table2}). On the other hand, we have another pair
of oscillators of the second kind $(V_{\rm sq}, m_{(n)})$ and
$(V_{\rm sq}, m_R)$, which are respectively isospectral with the
constant-mass potentials
\be
V_*(y_{(n)}) = \frac{\hbar \omega_0}{8} \left\{ \left[ \left(
\frac{2n+1}{\alpha y_{(n)}} \right)^{2n+1} - \left( \frac{\alpha
y_{(n)}}{2n+1} \right)^{2n+1} \right]^2 + 2(1-\sqrt{2})\right\},
\quad n\in\mathbb{N}
\label{pair2}
\ee
and
\be
V_*(y)=\frac{\hbar \omega_0}{8} \left\{ \left[ \frac{1}{x_0 +
\sinh \alpha y} - (x_0 + \sinh \alpha y) \right]^2
+2(1-\sqrt{2})\right\}.
\label{pair3}
\ee
Each one of these last potentials shows a spectrum which is a
distorted version of $\textrm{Sp} (V_{\rm sq}) =\textrm{Sp}
(V_{\rm osc})$. In summary, given an interaction represented by
$V(x)$, the spectrum of a position-dependent mass quantum system
is a distorted version of the spectrum of a particle of mass $m_0$
subject to the same interaction. As we have realized, the degree
of distortion depends directly on the explicit position dependence
of the involved mass.

\section{Factorization and Coherent States}
\label{sec4}

Once we have constructed the solvable position-dependent mass
Hamiltonians $H_a$, one can look for the appropriate factorization
operators. The presence of $m(x)$ in $H_a$ makes necessary a
refinement of the factorization (see e.g. \cite{Neg00b} and
\cite{Neg00}). As usual, the factorization operators intertwine
the initial Hamiltonian with a set of new exactly solvable
energy-like operators $\widetilde H_a$ \cite{Mie04,And04}.
However, in general they do not act as ladder operators on the
eigenfunctions of neither $H_a$ nor $\widetilde H_a$. In the case
of position-dependent mass oscillators of the first kind, the
factorization operators act in a ladder form if their commutator
is the appropriate constant. Then, as we are going to show, it can
be constructed a set of position-dependent mass coherent states.

\subsection{The position-dependent mass factorization}

Let $A$ and $B$ be the following operators
\be
A = -\frac{i}{\sqrt 2} m^a P m^b + \beta, \qquad B =
\frac{i}{\sqrt 2} m^b P m^a + \beta, \qquad A^{\dagger} = B,
\label{ab}
\ee
with $\beta$ a function of the position operator $X$. We want to
work with $A$ and $B$ as the factorization operators of $H_a$. In
this regard, it is important to stress that most of the literature
pay attention on a specific ordering of $m$ and $P$. Namely, it is
usual to take $a=0$ and $b=-1/2$ so that the kinetic part of $H_a$
reads $\frac12 P \frac{1}{m} P$, with the corresponding
simplification of $A$ and $B$ (see e.g. \cite{Mil97,Pla99,Roy05}).
Here, we shall use the operators (\ref{ab}) with no {\it a
priori\/} assumption on the ordering of $m$ and $P$. In this way,
the results already reported will be included as particular cases.

If $A$ and $B$ factorize the Hamiltonian (\ref{hamil1}) in a {\it
refined\/} way \cite{Neg00} then one has
\be
H_a = AB + \epsilon,
\label{factor1}
\ee
and $\beta$ fulfills a Riccati equation in the position
representation:
\be
V -\epsilon = \frac{\hbar}{\sqrt{2m}} \left[ 2\left( a +
\frac{1}{4} \right) \left( \frac{m'}{m} \right) \beta - \beta'
\right] + \beta^2
\label{riccati1}
\ee
where $\epsilon$ is a constant (in energy units) to be fixed. For
arbitrary $m$ and $\beta$ the product between the factorization
operators obeys the commutation rule:
\be
[A,B] = -\frac{\hbar^2}{m^3} \left(a+\frac{1}{4}\right) \left[
mm'' - \frac{3(m')^2}{2} \right] - \frac{2\hbar}{\sqrt{2m}}
\beta'.
\label{conmuta1}
\ee
Therefore we have a new operator $\widetilde{H}_a$, defined as
follows
\be
\widetilde{H}_a \equiv BA + \epsilon = K_a + \widetilde{V}, \qquad
\widetilde{V} := V -[A,B],
\label{hamilnew}
\ee
which is intertwined with $H_a$ by means of the factorization
operators:
\be
\widetilde H_a B = BH_a, \qquad H_a A = A\widetilde H_a.
\label{intertwin1}
\ee
The relevance of these last relationships is clear by noticing
that, if $\psi$ is an eigenfunction of $H_a$ with eigenvalue $E$
(see Eq. \ref{eigen1}), then $\widetilde \psi \propto B\psi \neq
0$ solves the new eigenvalue equation
\be
\widetilde H_a \widetilde \psi(x) = E \widetilde \psi(x).
\label{schro2}
\ee
Moreover, it is easy to verify that a normalized wavefunction
$\psi$ leads to $\vert \widetilde \psi \vert^2 \propto
E-\epsilon$. Then, the new set $\{ \widetilde \psi =
B\psi/(E-\epsilon)^{1/2} \, \vert \, E\neq \epsilon \}$ consists
of normalized eigenfunctions of $\widetilde H_a$ belonging to the
eigenvalues $\{ E\} = \textrm{Sp}(H_a)$. Now, let $\widetilde
\psi_{\epsilon}$ be a function which is orthogonal to the set $\{
\widetilde \psi \}$, i.e., $(\widetilde \psi, \widetilde
\psi_{\epsilon}) \propto (\psi, A \widetilde \psi_{\epsilon}) =0$.
Since $\psi \neq 0$ we have $A \widetilde \psi_{\epsilon} =0$ and
necessarily $\widetilde H_a \widetilde \psi_{\epsilon} = \epsilon
\widetilde \psi_{\epsilon}$. The involved solution reads
\be
\widetilde \psi_{\epsilon} = C_{\epsilon} \, m^{a+1/2} \exp
\left[\frac{\sqrt{2}}{\hbar} \int^x m^{1/2} \beta dr\right]
\label{missing1}
\ee
with $C_{\epsilon}$ a constant of integration. If $(\widetilde
\psi_{\epsilon},\widetilde \psi_{\epsilon})<\infty$ then ${\rm Sp}
(\widetilde H_a) = \textrm{Sp} (H_a) \cup \{\epsilon\}$.

The previous derivations considered $\epsilon \notin
\textrm{Sp}(H_a)$. To include the case $\epsilon \in {\rm
Sp}(H_a)$ let us assume that the solution of $B\psi_M =0$, given
by
\be
\psi_M = C_M C_{\epsilon} m^{1/2} (\widetilde
\psi_{\epsilon})^{-1},
\label{missing2}
\ee
is a square-integrable function. In this way $\psi_M$ is the
wavefunction of $H_a$ belonging to the eigenvalue $E=\epsilon$. As
a consequence, there is no element in $\{ \widetilde \psi \}$
constructed from $\psi_M$ via the relationships
(\ref{intertwin1}). The corresponding function $\widetilde \psi_M$
must be obtained as a solution of $BA \widetilde \psi_M=0$ (see
Eqs. \ref{hamilnew} and \ref{schro2}). There are two possible
cases:

1) If $A \widetilde \psi_M =0$, then $\widetilde \psi_M$ has the
same form as the function defined in (\ref{missing1}). However, if
$\psi_M \in L^2({\cal D}^{(a)})$, from Eq.~(\ref{missing2}) one
notices that $\widetilde \psi_M \propto m^{1/2}/\psi_M$ is not
square-integrable.

2) If $A \widetilde \psi_M \neq 0$ and $B(A \widetilde \psi_M)
=0$, one can take $A \widetilde \psi_M =\psi_M$ such that
$B(\psi_M)=0$. Then, because $\psi$ and $\psi_M$ are orthogonal,
we have $(\widetilde \psi, \widetilde \psi_M) = (\psi, A
\widetilde \psi_M) = (\psi,\psi_M)=0$. Thus, $\widetilde \psi$ and
$\widetilde \psi_M$ are orthogonal and $\textrm{Sp}(\widetilde
H_a) = \textrm{Sp}(H_a)$, with $\epsilon \in \textrm{Sp}(H_a)$.

At this stage it is important to stress that $\textrm{Sp} (H_a)$
and $\{ \psi \}$ can be obtained by means of the transformations
introduced in Section~\ref{sec2}. Thereby, we get a wide family of
isospectral operators if, for instance, $H^{(a_n)}$ is the
MDNT-Hamiltonian defined in (\ref{free}). That is, we have :
\be
\textrm{Sp}(\widetilde H_{a_n}) = \textrm{Sp}(H_{a_n})
\rightleftarrows \textrm{Sp} (H^{(a_n)}) = \textrm{Sp} (\widetilde
H^{(a_n)})
\label{espectros}
\ee
where $\widetilde H^{(a_n)}$ is the Hamiltonian intertwined with
$H^{(a_n)}$ in the $y_{(n)}$-representation. The same can be said
about the MINT-Hamiltonian $H$ defined in (\ref{free2}). In this
context, it will be profitable to decompose the commutator
(\ref{conmuta1}) in the MDNT and MINT cases:
\be
[A,B]= \left\{
\begin{array}{ll}
-\displaystyle\left[ \frac{\hbar}{m^{3/2}} \left( a+\frac{1}{4}
\right) m'\right]^2 -\displaystyle\sqrt{\frac{2\hbar^2}{m}}
\beta' & (\textrm{MDNT})\\[3ex]
 -\displaystyle\sqrt{\frac{2\hbar^2}{m}}  \beta' &
(\textrm{MINT})
\end{array}
\right.
\label{conmuta2}
\ee

\subsection{Position-dependent mass ladder operators}

Let us consider the Hamiltonian of an oscillator of the first kind
$H_a$. In advance we know that $\textrm{Sp}(H_a) = \{ E_n =\hbar
\omega_0(n+1/2) \}_{n=0}^{+\infty}$, whichever the MDNT or the
MINT case we are dealing with (see Section~\ref{sec3}). To get the
simplest form for the corresponding annihilation and creation
operators let us take $[A,B]=-\hbar \omega_0$. Then we have
$\widetilde H_a = H_a + \hbar \omega_0$. That is, $\widetilde H_a$
differs from $H_a$ only in the zero of the potential. This
physical equivalence and the intertwining relationships
(\ref{intertwin1}) make clear the roles played by the
factorization operators:
\be
A(H_a + \hbar \omega_0) = H_a A, \qquad B(H_a -\hbar \omega_0) =
H_aB.
\label{intertwin2}
\ee
Therefore $A\psi_n \propto \psi_{n+1}$ and $B\psi_n \propto
\psi_{n-1}$ if $H_a\psi_n = E_n \psi_n$. Now, the substitution of
$[A,B]=-\hbar \omega_0$ in (\ref{conmuta1}) leads to the following
$\beta$-function
\be
\beta = \frac{\omega_0}{\sqrt 2} \int^x m^{1/2} dr -
\frac{\hbar}{\sqrt 2} \left( a + \frac{1}{4} \right)
\left(\frac{m'}{m^{3/2}}\right) +
\beta_0.
\label{beta}
\ee
Here $\beta_0$ is an integration constant which will be omitted in
the sequel. The identification $\epsilon = \hbar \omega_0/2$,
after introducing (\ref{beta}) in the Riccati equation
(\ref{riccati1}), allows to write the potential $V$ in terms of
the $\beta$-function:
\be
V= \beta^2 + \frac{2\hbar}{\sqrt{2 m^3}} \left( a + \frac{1}{4}
\right) \left\{ m' \beta + \frac{\hbar}{2\sqrt{2 m^3}} \left[ mm''
- \frac{3}{2} (m')^2 \right] \right\}.
\label{pot}
\ee
The straightforward calculation shows that this last expression is
reduced in both the MDNT and the MINT cases to the same simple
form:
\be
V= \frac{\omega_0^2}{2} \left[ \int^x m^{1/2} dr \right]^2.
\label{pot2}
\ee
Remark that this expression for the potential is consistent with
our transformations in Sections~\ref{sec2} and \ref{sec3}. Indeed,
since $V_*(y)= \frac{m_0 \omega_0^2}{2} y^2$ has been given as the
initial potential, its $x$-representation reads
\be
V(x) = [V_*\circ s](x)=V_* (s(x)) = \frac{m_0 \omega_0^2}{2}
(s(x))^2
\label{pot3}
\ee
with $y=s(x)$ given in (\ref{gy}). In summary, we have shown that
$A$ and $B$, as they are defined in (\ref{ab}-\ref{riccati1}), are
nothing but creation and annihilation operators if their
commutator (\ref{conmuta1}) is constrained to be a constant equal
to the separation between the energy levels of $H_a$. The same
condition allows to identify a quadratic potential $V$, expressed
in terms of the mass-function, which is consistent with the
transformations defined in the previous sections.

\subsection{Position-dependent mass coherent states}

To take full advantage of the results derived in the previous
sections let us rewrite the factorization operators as follows
(compare with \cite{Mil97,Pla99,Roy05}):
\bea
A =-\frac{\hbar}{\sqrt{2m}} \left[ \frac{d}{dx} -\frac{(\ln
m)'}{4} \right] + \frac{\omega_0}{\sqrt 2} \int^x m^{1/2}
dr,\\[1ex]
B =\frac{\hbar}{\sqrt{2m}} \left[ \frac{d}{dx} -\frac{(\ln m)'}{4}
\right] + \frac{\omega_0}{\sqrt 2} \int^x m^{1/2} dr
\label{abnew}
\eea
where we have used (\ref{ab}) and (\ref{beta}). The operator $B$
in the $y$-space is then given by
\be
B_* = \frac{\hbar}{\sqrt{2 m_0}} \frac{d}{dy} + \left(
\frac{\omega_0^2 m_0}{2} \right)^{1/2} y - \frac{\hbar}{\sqrt{32
m_0}} \left( \frac{d \ln m_*}{dy} \right)
\label{bnew}
\ee
and a similar expression for $A_*$, obtained from (\ref{bnew}) by
changing the sign of the first and third terms. Finally, in the
dimensionless notation of Eq.~(\ref{free1}) we get
\be
\textrm{B}_* = \frac{d}{d\textrm{y}} + \textrm{y} - \left(
\frac{d}{d\textrm{y}} \ln m_*^{1/4}\right), \qquad B_* = \left(
\frac{\hbar \omega_0}{2} \right)^{1/2} \textrm{B}_*
\label{bfree}
\ee
and $A_* = \textrm{A}_*\sqrt{\hbar \omega_0/2}$. Then, from
(\ref{ab}) we have
\be
\textrm{H}_{a*} = (2/\hbar \omega_0) H_{a*} =\textrm{A}_*
\textrm{B}_* +1, \qquad [\textrm{A}_*, \textrm{B}_*] =-2
\label{newcon}
\ee
The action of $\textrm{H}_{a*}$, $\textrm{A}_*$ and $\textrm{B}_*$
on $\psi$ in the dimensionless $y$-representation is as follows:
\be
\textrm{H}_{a*} \psi_* = J_*^{1/2} (-\ddot \varphi + y^2 \varphi),
\qquad \textrm{A}_* \psi_* = J_*^{1/2} a_+ \varphi_*, \qquad
\textrm{B}_* \psi_* = J_*^{1/2} a_- \varphi_*.
\label{escala}
\ee
The Jacobian $J$ is defined in Eq. (\ref{map2}) and $a_-$ ($a_+$)
is the conventional annihilation (creation) operator of the linear
oscillator in the y-representation
\be
a_- := \frac{d}{d\textrm{y}} + \textrm{y}, \quad (a_+)^{\dagger}
=a_-, \qquad [a_-,a_+]=2, \qquad a_+ a_- = 2N
\label{aniq}
\ee
with $N$ the Fock's number operator. Hereafter we shall omit the
``$*$-notation''.

In order to construct a set of coherent states as eigenfunctions
of $\textrm{B}$ we first take an arbitrary linear combination
$\Theta$ of the wavefunctions $\psi_n$ associated with
$\textrm{H}_a$:
\be
\Theta = \sum_{k=0}^{\infty} c_k \psi_k.
\label{teta1}
\ee
The action of $\textrm{B}$ on this last function reads
\be
\textrm{B} \Theta = J^{1/2} \sum_{k=0}^{\infty} c_k \sqrt{2k} \,
\varphi_{k-1}.
\label{coh1}
\ee
We look for the functions $\Theta$ fulfilling $\textrm{B}
\Theta=z\Theta$, $z \in \mathbb{C}$. The straightforward
calculation leads to a recurrence relation which is satisfied by
the coefficients $c_k$. The root is found to be $c_k = z^k
c_0/\sqrt{k!2^k}$. As usual, the coefficient $c_0$ is fixed by the
normalization of $\Theta$ and we finally arrive at the familiar
expression:
\be
\Theta_z = e^{-\frac{\vert z\vert^2}{4}} \sum_{k=0}^{\infty}
\frac{z^k}{\sqrt{2^k k!}} \psi_k = J^{\frac12} e^{-\frac{\vert
z\vert^2}{4}} e^{\frac{za_+}{2}} \varphi_0 \equiv J^{\frac12}
e^{-\frac{\vert z\vert^2}{4}} e^{\frac{za_+}{2}}
e^{-\frac{\overline{z}a_-}{2}} \varphi_0 = J^{\frac12} D(z)
\varphi_0
\label{coh2}
\ee
where $\overline z$ stands for the complex conjugation of $z$ and
we have used the Baker-Campbell-Hausdorff formula $e^A e^B = \exp
(A+B+\frac12 [A,B])$ to recover the displacement operator
\be
D(z) := e^{\frac{za_+ -\overline{z} a_-}{2}} = e^{za_+/2}
e^{-\overline{z} a_-/2} e^{-\vert z\vert^2/4}.
\label{desplaza}
\ee
Thereby, since $\theta_z(y) := D(z) \varphi_0(y)$ is a
conventional constant-mass coherent state in the $y$-space we
conclude that its partner in the $x$-coordinates $\Theta_z =
J^{1/2} \theta_z$ is a position-dependent mass coherent state,
defined in terms of the annihilation operator $B$. Explicitly,
$\Theta_z(x)$ is given by
\be
\Theta_z(x) = \left( \frac{m(x)\, e^{-\vert z\vert^2}}{m_0}
\right)^{1/4} \sum_{k=0}^{\infty} \frac{z^k}{\sqrt{2^k k!}} \,
\varphi_k(s(x)).
\label{coherente}
\ee
This last result is the general expression for the coherent states
of any of the oscillators of the first kind introduced in
Section~\ref{sec31}. In this context, let one of these oscillators
be in the state $\Theta_z$. The probability of getting
$\textrm{E}_n= 2n+1$ as the result of a measurement of the energy
is ruled by the Poisson distribution:
\be
{\cal P}_n(\Theta_z) \equiv \vert (\psi_n, \Theta_z) \vert^2 =
\frac{\vert z\vert^{2n}}{2^n n!} \, e^{-\vert z \vert^2/2}.
\label{poisson}
\ee
The mean value $\langle \textrm{H}_a \rangle_z$ is then given by
\be
\langle \textrm{H}_a \rangle_z \equiv (\Theta_z, \textrm{H}_a
\Theta_z) = \sum_{k=0}^{\infty} {\cal P}_k (\Theta_z) \textrm{E}_k
= \vert z \vert^2 +1
\label{mean}
\ee
where we have used Eq. (\ref{newcon}). In the same manner we find
$\langle \textrm{H}_a^2 \rangle_z = \vert z \vert^4 +4\vert z
\vert^2 +1$, so that $\Delta \textrm{H}_a = \vert z \vert
\sqrt{2}$. Hence, for very large $\vert z \vert$ one gets $\Delta
\textrm{H}_a << \langle \textrm{H}_a \rangle_z$ and the relative
value of the energy of the state $\Theta_z$ is well defined, as
usual for the Glauber states. It is also simple to verify that the
product of the root-mean-square deviations $\Delta P$ and $\Delta
Y$ is minimized. In conclusion, the $z$-parameterized functions
(\ref{coherente}) are the coherent states belonging to a wide
class of position-dependent mass oscillators of the first kind.

\section{Concluding remarks}

We have studied the problem of solving the Schr\"odinger equation
for an arbitrary position-dependent mass system. Our approach is
useful to face two general physical situations. In the first one
we look for the interaction which must be applied on a mass $m(x)$
to supply it with a particular spectrum of energies. The second
physical situation corresponds to the case in which one is
interested on a given position-dependent mass $m(x)$, subject to a
particular interaction $V(x)$ rather than in the recovering of a
specific spectrum. For arbitrary orderings of $m(x)$ and $P$ in
the Hamiltonian, diverse general expressions for $m(x)$ were
derived as a consequence of mapping the original Schr\"odinger
equation to a conventional constant mass one. It was also found
that the transformation is rather simple for the very special
ordering $m^{-1/4}P m^{-1/2} P m^{-1/4}$ in the kinetic part of
the Hamiltonian. In contradistinction with
\cite{Mil97,Pla99,Roy05}, we showed that a position-dependent mass
Hamiltonian can be factorized as the product of two mutually
adjoint operators with no a priori assumptions on the ordering of
$m$ and $P$.

In particular, two kinds of position-dependent mass oscillators
were analyzed. The first one is defined to be isospectral with the
quantum oscillator of mass $m_0$, no matter the explicit form of
$m(x)$ and $V(x)$. The oscillators of the second kind exhibit
spectra different from the equidistant energies $\hbar
\omega_0(n+1/2)$ and correspond to a particular mass $m(x)$
subjected to the harmonic oscillator potential. Results include
the singular oscillator as well as confining odd-root-law, $\ln^2$
and $\sinh^2$-like interactions. The special case of a particle of
mass $m_0$ in a confining even-power-law potential was found to be
isospectral to a system of mass $m(x) \propto x^{-4n/(2n+1)}$,
subject to the action of the harmonic oscillator potential. The
factorization operators were then selected to work as ladders in
the space of the position-dependent mass oscillators. Finally, the
coherent states corresponding to oscillators of the first kind
were explicitly constructed as eigenvectors of the annihilation
operator. These new CS have the same analytical form as the
Glauber states and minimize the position-momentum uncertainty
principle.

Of special interest, the singular oscillator $V_{\rm sq}$ defined
in Eq. (\ref{spec1}) and referred in Section~\ref{seceqz} as the
{\it squeezed oscillator\/}, exhibits CS connected with the
$su(1,1)$ Lie algebra if the mass is a constant \cite{Per86}.
Particular cases of the mass-function have been shown to preserve
the $su(1,1)$ spectrum-structure of $V_{\rm sq}$ \cite{Roy05}. The
same is true for any of the masses derived in this paper. Thereby,
it is sound to construct position-dependent mass $su(1,1)$-like
CS. Following \cite{Per86}, such result could be applied to get a
better understanding of the physics of $N$ interacting particles
(work in this direction will be published elsewhere \cite{Cru09}).
Other physically interesting systems can be analyzed in the
corresponding manner once the dynamical algebra is given. Special
attention must be drawn to the Susy non-linear algebras engaged
with infinite point spectra. If the energy levels can be obtained
by a function of its index $E_n =E(n)$ then one can distinguish
between {\it natural\/} and {\it linear\/} algebras of the
Susy-system. To each one of these algebras there exists a
companion set of CS \cite{Fer07}. Then, besides the systems above
discussed, it would be also interesting to analyze the
position-dependent mass CS belonging to higher-order Susy
potentials like the P\"oschl-Teller ones \cite{Dia99} and
\cite{Fer08} (see also \cite{Cru08}). Results on the matter are in
progress.

\section*{Acknownledgementes}
The support of CONACyT project 24333-50766-F and IPN grants COFAA
and EDI is acknowledged. ORO is grateful to Y. Gurevich for
enlightening comments. SCyC thanks the members of Physics
Department, Cinvestav for kind hospitality.


\end{document}